\documentclass[12pt]{iopart}

\usepackage{graphicx}
\usepackage{deluxetable}
\usepackage{wrapfig}
\usepackage{sidecap}
\begin{document}

\title[NANOGrav]{The North American Nanohertz Observatory for Gravitational Waves}

\author{M. A. McLaughlin}

\address{Department of Physics, West Virginia University, Morgantown, WV 26506-6315, USA}
\ead{maura.mclaughlin@mail.wvu.edu}
\begin{abstract}
The North American Nanohertz Observatory for Gravitational Waves (NANOGrav)
is a collaboration of researchers who are actively engaged in using North American radio telescopes to detect and study gravitational waves via pulsar timing.
To achieve this goal, we regularly observe millisecond pulsars (MSPs) with the Arecibo and Green Bank Telescopes and develop and implement new instrumentation
and algorithms for searching for and observing pulsars, calculating arrival times, understanding and correcting for propagation delays and sources of noise in our data, and detecting and characterizing a variety of gravitational wave sources. We collaborate on these activities with colleagues in the International Pulsar Timing Array (IPTA). We also educate students of all levels and the
 public about the detection and study of gravitational waves via pulsar timing.
\end{abstract}

\section{Introduction}

Gravitational waves (GWs), ripples in space-time produced by accelerating massive objects, are a fundamental prediction of Einstein's theory of general
relativity. Measurements of orbital decay due to GW emission in double neutron star binary systems provide convincing evidence for their existence \cite{tw82,ksm+06}. However,
as of yet, we have not detected the influence of GWs on space-time through a measured change in light travel time between two objects. This direct detection of GWs will
immediately provide spectacular proof of Einstein's theories and will also usher in a new era of astronomy in which we can use GWs to study objects which are thus far invisible or
inaccessible through electromagnetic observations.

Pulsars are rapidly rotating, highly magnetized neutron stars produced in the supernova explosions of massive stars.
Pulsar timing arrays (PTAs) are able to detect GWs through high-precision timing, sensitive to small changes in the light travel
times between the pulsars and Earth. The pulsars used for our experiment are millisecond pulsars (MSPs), which have been spun-up to very short periods through accretion of mass and angular momentum from a companion star. These objects are incredibly stable rotators, with arrival times measurable to microsecond precision and spin periods predictable to one part in $10^{15}$.
There are over 200 known Galactic MSPs, of which approximately 50 are being observed by PTAs using the largest radio telescopes in the world, including two in North America, one in Australia, and five in Europe.

PTAs are sensitive to GWs with periods comparable to the total time span of our observations. Given PTA experiments of 5--10 year durations,
our sensitivity will peak in the
 $10^{-8}-10^{-9}$~Hz frequency range, complementary to the much higher frequencies probed by ground- or spaced-based
GW interferometers. 
The most likely GW sources for detection by PTAs include supermassive black hole binaries,  cosmic strings, and, possibly, early universe inflation. Therefore, PTAs will provide crucial input to galaxy formation and evolution scenarios and cosmology.
Additional source classes also may await discovery.

NANOGrav was formed in October 2007 as a collaboration of researchers at North American universities, colleges, national laboratories, and observatories. We use  the 100-m Green Bank Telescope (GBT) in Green Bank, WV, and the 300-m Arecibo Observatory (AO) in Arecibo, Puerto Rico to observe an array of MSPs with the goal of GW detection. It is one of three PTA
collaborations and, along with the European Pulsar Timing Array (EPTA) and the Parkes Pulsar Timing Array (PPTA), is a member of the International Pulsar Timing Array (IPTA).
In this article, we will review the collaboration's organization and then provide an overview of NANOGrav's activity in key areas.

\section{Organization}

\begin{figure}
\includegraphics[scale=0.4]{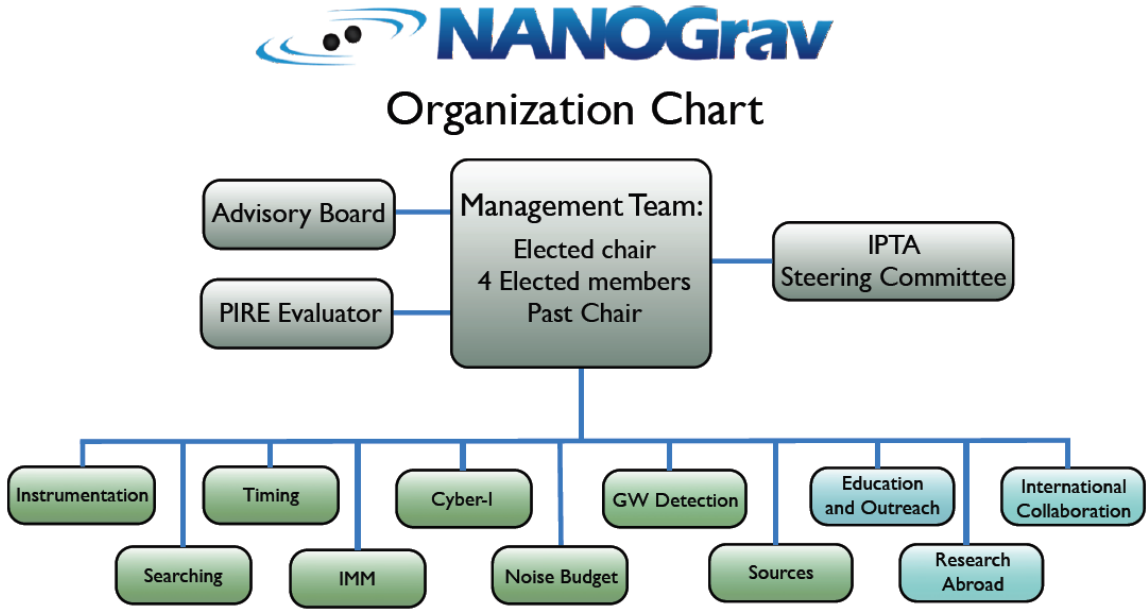}
\caption{NANOGrav is governed by a management team (MT) consisting of the Chair, four elected members, and (for one year after the end of their term) the previous chair. A six-member external Advisory Board evaluates the collaboration's activities and offers recommendations twice yearly.    The PIRE evaluator (see \S\ref{sec:pire}) provides specific input to the MT on PIRE-supported activities. The MT liases with the IPTA Steering Committee through two representatives, one of whom is also on the MT. We accomplish our goals through a number of science working groups, including Instrumentation, Timing, Cyber-Infrastructure, GW Detection, Searching, Interstellar Medium Mitigation (IMM), Noise Budget, and Sources. Other critical parts of NANOGrav work  include Education and Outreach, International Collaboration, and Research Abroad. These aspects are currently supported through our PIRE program.}
\label{fig:org}
\end{figure}

The organizational structure of NANOGrav is illustrated in Figure~\ref{fig:org}.
NANOGrav consists of faculty, senior researchers, postdocs, graduate students, and undergraduate students at 13 institutions in the United States and Canada. Membership is open to all 
who
share our goal of GW detection and study using pulsars. We currently have 31 Full,
14 Associate, and 12 Junior (or undergraduate) members.
New membership requests can be submitted to the Chair at any time and are accepted from participants in any country who are willing to contribute to our goals. To be eligible for Full Membership, a participant must be in the collaboration for
at least one year.
Our authorship policy states that all Full Members of NANOGrav will be authors on detection or upper limit papers.
NANOGrav is governed by a Chair and a management team. The Chair and the management team (MT) members are elected for two-year terms. Two NANOGrav members also serve on the IPTA Steering Committee. More details about our by-laws, membership policy, and authorship policy can be found at http://nanograv.org/governance.html.

The detection and study of GWs using pulsars requires a diverse portfolio of work in various observational, analytical, and theoretical areas. NANOGrav
accomplishes its goals in these areas through
various working groups. We being by describing the work involved in observing and timing an array of MSPs and in applying algorithms to detect and characterize sources, as these are the core activities of NANOGrav. We then describe how 
searching for MSPs, developing techniques to mitigate interstellar medium effects, and characterizing the pulsar noise budget are
 increasing our sensitivity to GWs.

\section{Timing Observations and Analysis}
\label{sec:timing}

\begin{SCfigure}
\includegraphics[scale=0.5]{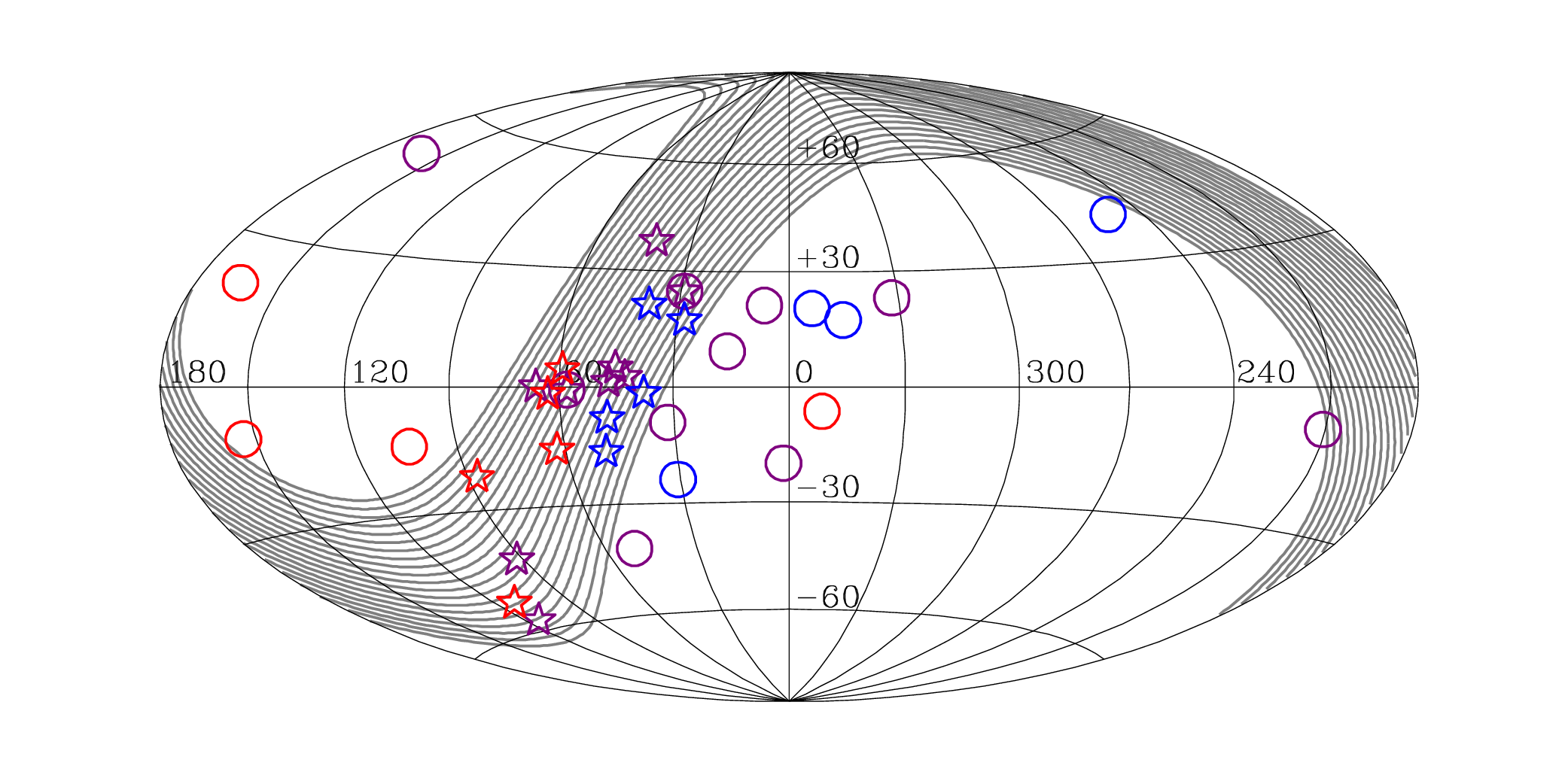}
\caption{The pulsars being timed by NANOGrav. The stars/circles denote MSPs timed with Arecibo/GBT. Objects in purple have been observed since 2005 or earlier and objects in blue/red since 2009/2011. Lines indicate the region of the sky visible to Arecibo.
  Figure credit: David Nice.}
\label{fig:aitoff}
\end{SCfigure}

Gravitational wave detection requires observations of a number of pulsars with the highest precisions possible.
A detailed discussion of timing methodology and algorithms is given by Demorest \& Lommen in this issue.
We have observed 18 pulsars for more than five years (see Figure~\ref{fig:aitoff}), with 17 of them used in the analysis for
NANOGrav's first paper reporting an upper limit on the GW background \cite{dfg+13}. Over the past five years,  three to four pulsars per
year have been added to the timing program  and a total of 36 pulsars are currently being observed. In this section, we first describe the timing observations used for the first upper limit paper and then the current observing scheme.

Preserving the narrow intrinsic pulse profiles which provide the highest timing precisions requires coherent dedispersion, or the removal of dispersive delays (see \S\ref{sec:imm}) by convolving
 the raw signal voltage with the inverse of the interstellar medium (ISM) transfer function \cite{lk05}.
The data used for NANOGrav's first upper limit paper were obtained between 2005 and 2010 with the previous generation coherent dedispersion backends  ASP (Arecibo Signal Processor)
and GASP (Green Bank ASP). The maximum bandwidth used was 64 MHz and each pulsar was observed at multiple frequencies for 15--45 minutes per integration and 1-minute (for Arecibo) or 3-minute (for GBT) sub-integrations. Two widely separated radio frequencies were used to allow for the correction of frequency-dependent
 interstellar delays
 (see \S\ref{sec:imm}). As neither telescope has a dual-frequency receiver, these observations were not simultaneous but occurred
on the same day at Arecibo and within several days of each other  at the GBT. Data were dedispersed within 4-MHz subbands. Each pulsar was observed roughly once every four to six weeks at each frequency. The
timing analysis consisted of cross-correlating profiles with noise-free templates to measure pulse times-of-arrival (TOAs). Dispersion measures (DMs) were fitted for each pulsar at
each epoch (see \S\ref{sec:imm}). For each pulsar, a constant time offset for each 4-MHz frequency channel was fit to account for pulse
shape evolution with frequency.
As shown in Table~\ref{tab:pulsars}, the 17 pulsars
have root-mean-square (RMS) residuals ranging from 30~ns (for PSR~J1713+0747) to 1.5~$\mu$s (for PSR~J1643$-$1224).
Red  noise  (see \S\ref{sec:noise}) was detected with high significance in the residuals of  two objects and with marginal significance in the residuals of two others. The remainder of the pulsars have residuals
consistent with white noise.

Our current  timing program consists of regular observations of 36 pulsars with Arecibo and the GBT   
  (see Table~\ref{tab:pulsars}). Observations are carried out roughly once every three weeks at both telescopes, with
each pulsar observed for about 10--40 minutes at two widely separated radio frequencies.
At the GBT, data are
accumulated and coherently dedispersed in 1.5625-MHz wide subbands using the Green Bank Ultimate Pulsar Processing Instrument (GUPPI), an FPGA-based spectrometer  capable of processing  up to 800 MHz of bandwidth \cite{drd+08}.  A bandwidth of 200~MHz is used for observations at a center frequency  of 820~MHz and a bandwidth of 800 MHz is used for observations at a center frequency of  1500~MHz. At Arecibo, the PUPPI backend coherently dedisperses data, also in 1.5625-MHz wide subbands, over bandwidths of 40/700/600 MHz at center frequencies of 430/1410/2310 MHz, respectively. Data are folded in real-time with 15-s integrations using GUPPI and 10-s subintegrations using PUPPI. A calibration scan is obtained for each pulsar
by injecting a 25-Hz noise diode into the signal path for both polarizations.
At each observation epoch and at each frequency, a flux calibrator (B1442+101) is observed.

\begin{SCfigure}
\includegraphics[scale=0.5]{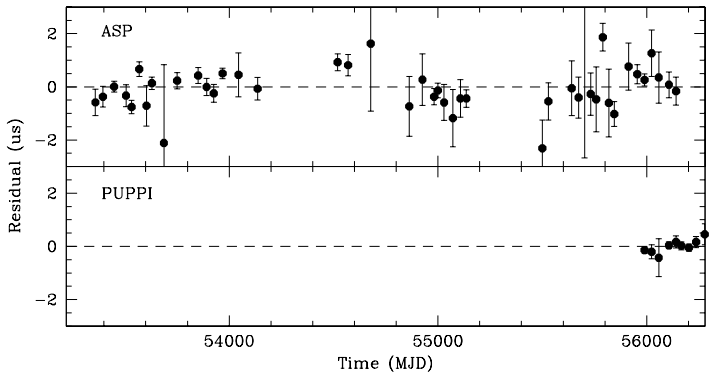}
\caption{Residuals for PSR~J0030+0451 at 1400~MHz with Arecibo using the ASP and PUPPI backends. The RMS residual has decreased by roughly the factor of three expected due to the increase in  bandwidth.}
\label{fig:0030}
\end{SCfigure}

Some properties of the pulsars in our current sample and some observational details  are provided in Table~\ref{tab:pulsars}.
A ``living'' version of this table, with  updated RMS values and median TOA uncertainties, is available at http://nanograv.org/sources.html.
Naively, the over 10-fold increase in bandwidth due to GUPPI and PUPPI should result in arrival times measured with over three-fold increased precision at some frequencies, leading to RMS residuals over three times smaller than those listed in Table~\ref{tab:pulsars}. Actual improvements (see Figure~\ref{fig:0030} for an example) in RMS residual range from factors of $\sim$1 (i.e. no improvement) to 3.5 (i.e. expected improvement). Smaller improvements may indicate that some 
 residuals are dominated by pulse amplitude or phase jitter, ISM effects, or, possibly, red spin noise. We are currently implementing new methods
for TOA calculation over large bandwidths which fit for both  DM and pulse phase offset simultaneously \cite{pennucci}. 
Using these techniques on the sensitive GUPPI and PUPPI data will result in a new timing data release and subsequent correlation analysis,
with a likely significant improvement to the current stochastic background upper limit.

In addition to analyzing data with the new backends, we are extending our data span back in time through the inclusion of
 archival data. Roughly 20 years of timing data on PSR~J1713+0747 are being analyzed to calculate what will likely be the most sensitive 
single-pulsar GW upper limit  \cite{zhu13}. In addition,
 data taken with the Wideband Arecibo Pulsar Processors (WAPPs) over 300-MHz of bandwidth at the same time as the ASP data are being used to improve the current limit from correlation analysis and offer insights into the sensitivity trade-off between coherent dedispersion and larger non-coherently dedispersed bandwidths  \cite{swiggum13}.

\begin{table}[h!]
\caption{Pulsars observed by NANOGrav}
\footnotesize\rm
\begin{tabular}{cccccccccc}
\hline
Name & Period & DM & Obs. & Frequencies & RMS$_{\rm ASP}$ & $\tilde{\sigma_{t,\rm ASP}}$ &MJD & T & N \\
PSR & (ms) & (pc~cm$^{-3}$) & & (MHz) & ($\mu$s) & ($\mu$s) & \\
\br
J0023+0923  & 3.05 &  14.3 & AO & 430/1410& -- & -- &  55731  & 1.9 & 56\\
J0030+0451 & 4.87  & 4.3 &AO& 430/1410& 0.148 & 0.37 &  50788 & 15.5 & 325\\
J0340+4130  & 3.30  & 49.6 & GBT & 820/1500 & -- & -- & 55972 & 1.3 & 41\\
J0613$-$0200 & 3.06 & 38.8 & GBT & 820/1500& 0.178 & 0.30 & 53348 & 8.4 & 217 \\
J0645+5158 & 8.85 & 18.2 & GBT & 820/1500& -- & -- & 55700  & 2.0 & 53\\
J0931$-$1902 & 4.64 & 41.5 & GBT & 820/1500& -- & -- & 56351 & 0.2 &53\\
J1012+5307 & 5.26 & 9.0 &  GBT & 820/1500&0.276 & 0.67 & 51012  & 14.8 & 236 \\
J1024$-$0719 & 5.16 & 6.5 & GBT& 820/1500&-- & --& 55094 & 3.7 & 120 \\
J1455$-$3330 & 7.99 & 13.6 & GBT& 820/1500&0.787 & 2.35& 53217  & 8.8 & 232 \\
J1600$-$3053 & 3.60 & 52.3 &  GBT& 820/1500&0.163& 0.34& 54400  & 5.6& 173\\
J1614$-$2230 & 3.15 & 34.5 & GBT & 820/1500&--&--& 54724  & 4.7 & 134\\
J1640+2224 & 4.62 & 62.4 & AO &430/1410&0.409&0.22& 50788 & 15.5 & 279\\
J1643$-$1224 & 4.62 & 62.4 & GBT& 820/1500&1.467&0.67& 53217  & 8.8 & 236 \\
J1713+0747 & 4.57 & 16.0 &  AO & 1410/2030&0.030&0.08& 48738  & 21.2 &  659 \\
& & & GBT & 820/1500 & --  & -- & 53798  &  7.2 &  212 \\
J1738+0333 &  5.85 &  33.8 & AO &1410/2030& --&--& 54999  & 3.9 & 79 \\
J1741+1351 &  3.75&  24.0& AO& 430/1410& --&--& 54998  & 4.0 & 100 \\
J1744$-$1134 & 4.07 & 3.1 & GBT&820/1500&0.198&0.22& 53216  & 8.8 & 226 \\
J1747$-$4036 & 1.65 & 152.9 & GBT&820/1500&--&--& 55976  & 1.2 & 41 \\
J1853+1303  &4.09  &30.6 & AO & 430/1410& 0.255&0.61& 53370  & 8.4 & 126 \\
B1855+09 & 5.36&  13.3 & AO& 430/1410& 0.111&0.41& 46436  & 27.4 & 853 \\
J1903+0327  &2.15& 297.5 & AO& 1410/2030&--&--& 54357  & 5.7  & 82 \\
J1909$-$3744 & 2.95 & 10.4 &  GBT& 820/1500&0.038&0.09& 53219 & 8.8 & 212 \\
J1910+1256 &4.98  &38.1 &AO &1410/2030&0.708&0.40& 53370  & 8.4 & 127 \\
J1918$-$0642 & 7.65 & 26.6 & GBT& 820/1500&0.203&0.62& 53216 & 8.8 & 218 \\
J1923+2515 & 3.78 & 18.9 &AO &430/1410&--&--& 55493  & 2.6 & 62 \\
B1937+21 & 1.56 & 71.0 & AO & 1410/2030&--&--& 45985 & 28.6 &  1001 \\
& & & GBT & 820/1500&--&--& 53216  & 8.8 &  233 \\
J1944+0907 & 5.19 & 24.3 & AO & 430/1410& --&--& 54505  & 5.3 &  103 \\
J1949+3106 &13.14 & 164.1 & AO & 1410/2030&--&--& 51949 & 12.3 &  29 \\
B1953+29 & 6.13 & 104.5 & AO & 430/1410&1.437&1.49& 46112 & 28.2 & 144 \\
J2010$-$1323 & 5.22 & 22.2 &  GBT& 820/1500& --&--& 54725  & 4.7 & 125 \\
J2017+0603  &2.90 &  23.9 &AO & 1410/2030&--&--& 55500  & 2.6 &  69 \\
J2043+1711 &  2.38&  20.7 & AO & 430/1410&--&--& 55731  & 1.9 & 85\\
J2145$-$0750 & 16.05 & 9.0 &  GBT & 820/1500&0.202&0.57& 53219  & 8.8 & 187\\
J2214+3000  &3.12 &  22.6 &AO & 1410/2030&--&--&52214  & 11.6 & 69 \\
J2302+4442&  5.19&  13.8 &GBT & 820/1500&--&--& 55972  & 1.3 & 39 \\
J2317+1439 &  3.45&  21.9 &AO & 430/1410&0.251&0.19&48862  & 20.7 & 406\\
\end{tabular}
\tablecomments{Pulsar name, spin period, DM, telescope at which the source is timed, center frequencies used for observation, residual RMS and median TOA uncertainty 
$\sigma_t$ from the first  NANOGrav upper limit paper using the ASP and/or GASP instruments \cite{dfg+13}, MJD at which the timing program began, total time span of the observations, and number of observing epochs. The RMS and median TOA uncertainty for PSR~J1713+0747 are for the combined GBT and Arecibo dataset.
}
\label{tab:pulsars}
\end{table}

\section{Detection and Characterization}
\label{sec:detection}

NANOGrav develops algorithms for detection of  continuous, burst, and stochastic GW sources. Continuous sources  emit at a roughly constant GW frequency.
Super-massive black hole binaries are the most likely source of continuous GWs for PTAs.  The review article by Ellis in this issue
describes a Bayesian analysis pipeline that will detect and characterize such sources. 
This builds on earlier work which presented a likelihood analysis in the time domain \cite{esc12} and comparisons of matched filtering
and power spectral summing methods for continuous sources \cite{ejm12}. 

Burst sources have signal durations much shorter than the total time spans over which we have been observing
and could be due to mergers of supermassive black holes (SMBHs),
 periastron
passages of compact objects orbiting a SMBH, or cusps on cosmic strings \cite{dv00}.  Bayesian pipelines have been constructed to detect burst sources, even when
the waveform cannot be determined or the source localized \cite{fl10}. Recent work shows that because burst signals grow with data span, 
  red noise can hinder the detection of bursts and, likewise, bursts could make the stochastic GW background more difficult to detect \cite{cj12}.

Stochastic sources are characterized by 
GWs that are not resolvable into individual sources. The stochastic background at our frequencies of interest likely includes contributions from supermassive black hole binaries and, possibly, cosmic strings or relic GWs from the early universe.
Algorithms for detection rely on measuring the correlation between the residuals of pairs of pulsars as a function of angular
separation to search
for the characteristic quadrupolar signature expected for an isotropic GW background \cite{hd83}. Anisotropic backgrounds or individual sources could also be detected through this method, as described in the Cornish \& Sesana review in this issue. The first NANOGrav upper limit paper calculated a 2-$\sigma$ upper limit on characteristic strain at a frequency of 1/yr of $7\times10^{-15}$ by computing covariance matrices of the post-fit residuals. In this work, the GW analysis was performed separately from the timing fit, but the properties of the timing fit were used to determine the amount of GW that may have been absorbed in the fit. A single-pulsar upper limit of $1.1\times10^{-14}$ was also presented,
based on the residuals of PSR~J1713+0747. Current work involves developing methods 
to calculate and interpret the optimal cross-correlation statistic \cite{abc+09} and more efficient, but approximate, maximum likelihood approaches \cite{esv13}. NANOGrav aims to make detection pipelines for all source classes available at http://nanosoft.sourceforge.net after development
and testing.

The article by Siemens et al. in this issue discusses the time-to-detection for a stochastic background of GWs under various assumptions.
They show that we are approaching the strong-signal regime, where the GW power is larger than the white noise, and that in that regime 
 the time-to-detection depends only weakly
on the cadence of observations and the white-noise RMS, and much more strongly on the number of pulsars. This is independent of any assumptions about red noise (though in the case of significant red  noise, adding more pulsars to the array becomes even more important). Using realistic simulations assuming NANOGrav's current observing program and modest improvements, they show that a detection could occur as early as 2016 and will occur by 2023 given reasonable assumptions about the expected GW amplitude range of the SMBH binary background.

It is important to note that our goal encompasses not only  detection of GWs, but also characterization of sources and GW astrophysics. Models for
 supermassive black hole binary populations depend
 critically on assumptions about early universe galaxy formation and evolution. Hence, our stochastic background upper limits are already providing useful input to the broader astronomical community. 
Once we make a detection of the stochastic background, we will be able to measure its
amplitude and, eventually, spectrum in order to discern different source contributions and further constrain those populations.
Detections of single sources will immediately provide an estimate of the binary period, and, possibly, eccentricity.
While the localization of single sources may initially be too poor to  allow for electromagnetic follow-up, this will improve with time as more pulsars with precisely determined distances are added to the array. Furthermore, we are working with the electromagnetic community to perform targeted searches for specific sources in NANOGrav data. These can provide astrophysical constraints on individual source properties \cite{jllw04}.
More details about the possibilities of joint GW/electromagnetic observations may be found in the review by Burke-Spolaor in this issue.

\section{Pulsar Searching}

The sensitivity of a PTA increases with the number of MSPs included in the array, making searches for MSPs extremely important to our mission. NANOGrav's searching working group provides support for
 multiple searches with both Arecibo and the GBT.
Two pulsars searches are underway using Arecibo: the 327-MHz Arecibo Drift-Scan survey (AO327) and the 1.4-GHz Arecibo L-band Feed Array Survey (PALFA). AO327 is sensitive to nearby MSPs out of the
Galactic plane, while PALFA is sensitive to distant MSPs in the plane. 
Thus far, AO327  has discovered 20 pulsars, including two MSPs \cite{dsb+13}, and PALFA has discovered 116 pulsars, including 17 MSPs \cite{csl+12,dfc+12}. Note that only a fraction of MSPs found in a particular survey will have the narrow, bright profile and timing stability essential for inclusion in a PTA. Of the PALFA MSPs, two have thus far have been included in NANOGrav's regular timing program. 

Recent GBT searches at 350 MHz include the GBT Drift-Scan Survey (GBTDrift) and the Green Bank Northern Celestial
Cap Survey (GBNCC). GBTDrift discovered 35 pulsars, including seven MSPs \cite{blr+13,lbr+13}, and GBNCC has so far discovered 62 pulsars, including nine MSPs. One pulsar from GBTDrift and one pulsar from GBNCC have already been included in NANOGrav's timing program. One additional MSP was discovered
in a separate portion of the GBTDrift survey set aside for analysis by high-school students in the Pulsar Search Collaboratory program \cite{rsm+13}. In addition, 28 MSPs have been discovered
through GBT searches of unidentified {\it Fermi} sources \cite{rrc+13,bcc+13}, with three of these pulsars included in NANOGrav's timing program.

\begin{SCfigure}
\includegraphics[scale=0.5]{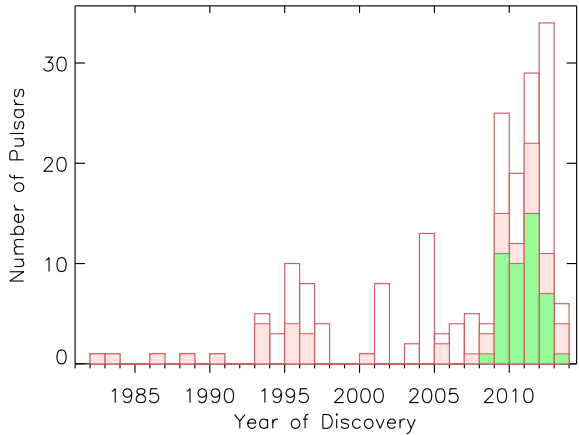}
\caption{The number of known Galactic MSPs\footnote{http://astro.phys.wvu.edu/GalacticMSPs} vs. year, with discoveries by the GBT in green and Arecibo in red. Over 60\% of all Galactic MSPs have been discovered since 2009. The GBT and Arecibo together have discovered roughly half of all Galactic MSPs. }
\label{fig:searches}
\end{SCfigure}

\begin{figure}
\includegraphics[scale=0.5]{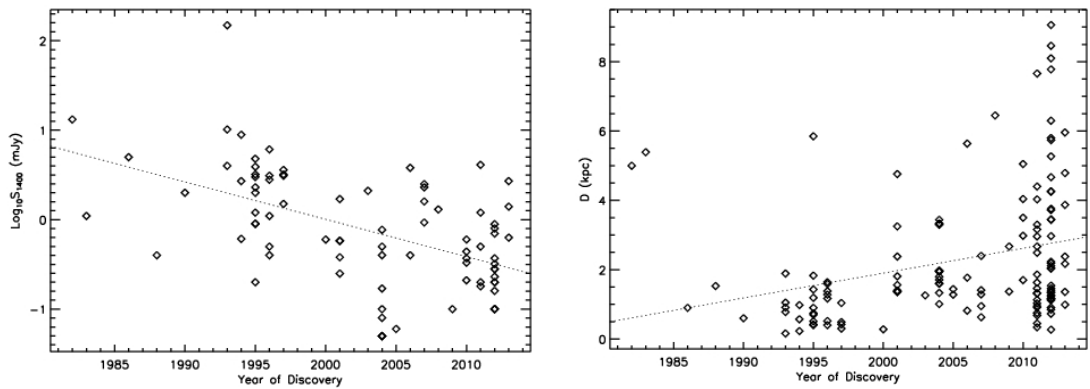}
\caption{Flux density at 1400 MHz (left) and distance (right) vs. year of discovery for Galactic MSPs$^\dagger$. Recent surveys continue to reveal bright, nearby MSPs.}
\label{fig:mspsurveys}
\end{figure}

 Figure~\ref{fig:searches} illustrates the rate of MSP discovery since the discovery of the first MSP in
1982, and the importance of the GBT and Arecibo surveys in which NANOGrav members are involved. 
The review by Stovall, Lorimer, \& Lynch in this issue offers more details about these surveys and
projections for the future. Recent MSP population studies show that there are a large number (roughly 30,000--80,000) of Galactic MSPs which remain to be detected \cite{lor13,lbb+13}. The precision with which we can time a pulsar is directly proportional to its flux density; current surveys are still detecting bright MSPs, demonstrating that continued searches may yield rich returns (see Fig.~\ref{fig:mspsurveys}). In addition,
current surveys continue to reveal nearby MSPs for which we are more likely to be able to measure precise distances through radio
interferometry (see Fig.~\ref{fig:mspsurveys}).

\section{Interstellar Medium Mitigation}
\label{sec:imm}

 Two frequency-dependent ISM effects - dispersion and scattering -  affect pulse TOAs. These are discussed in detail in the review by Stinebring in this issue. Dispersion due to refraction by free electrons results in delays proportional to DM  $\times \nu^{-2}$, 
 where the DM is the integrated column density of electrons along the line of sight and $\nu$ is radio frequency.
 Observations at widely separated frequencies (see Table~\ref{tab:pulsars}) maximize the differential dispersive delays and allow accurate DM fitting. Because of the relative motion
between pulsars and the ISM, DMs can change on timescales of $\sim$weeks or less, necessitating that the  observations at different frequencies
occur within, at maximum, one week  of each other.

The second frequency-dependent effect is scattering due to multipath propagation through the ISM. This results in
broadened pulse profiles, with a characteristic exponential tail and time delays 
 roughly proportional to $\nu^{-4}$.  Scattering delays for some MSPs can be greater than several $\mu$s at our observing frequencies and correction is crucial for achieve the highest  precisions possible. 
 NANOGrav is exploring several methods for removal of scattering delays. The first involves estimating the characteristic
bandwidth  $\Delta\nu_d$ through auto-correlation of dynamic
spectra, which describe how the pulsar flux changes with time and frequency.
 Then, the pulse broadening time, which is equal to the scattering delay under certain assumptions \cite{cr98}, can be calculated as
$(2\pi\Delta\nu_d)^{-1}$. This method may result in a modest improvement in RMS for some pulsars \cite{levin}.

The second method is termed cyclic spectroscopy (CS) and relies on the periodic signature of the pulsar signal to use the phase information to recover unscattered pulse  profiles and scattering delays \cite{dem11}. We are exploring the use of CS to calculate delays and correct TOAs using both simulated and real data. We are also applying the method to
MSPs with a variety of fluxes, pulse shapes, and DMs to determine the range of applicability as
the only published CS application is on a very bright, moderately scattered pulsar \cite{dem11}.
 Because baseband-sampled data is necessary,
we are  developing a GPU-based real-time CS implementation that can be applied to
every NANOGrav observation \cite{jones13}.

\section{Noise Budget}
\label{sec:noise}

\begin{figure}
\includegraphics[scale=0.6]{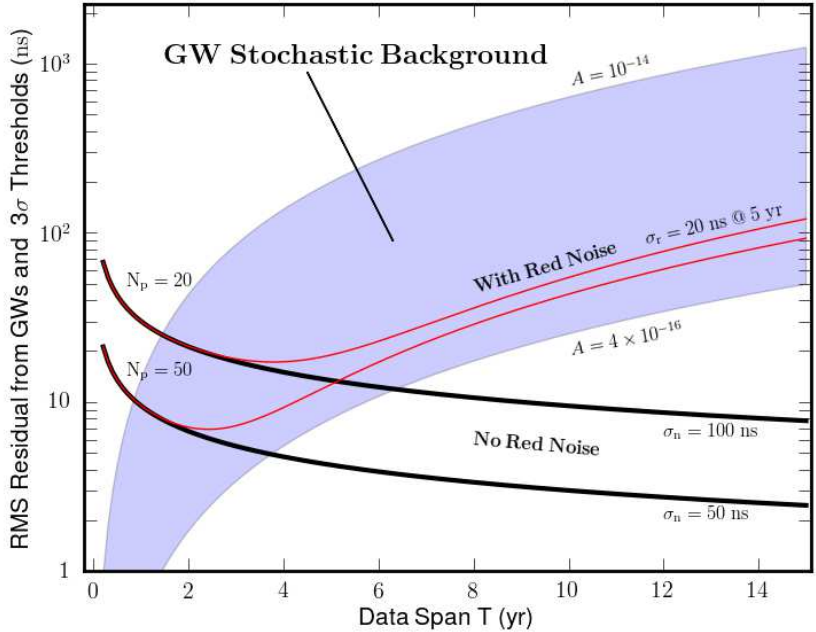}
\caption{The RMS timing perturbation from the GW stochastic background vs the data span $T$ (shaded region)  for a range of plausible values for the GW amplitude
   RMS due to supermassive black hole binaries. 
   The GW spectrum is $\propto f^{-13/3}$, while red spin noise is $\propto f^{-5}$, though as discussed in the review by Cordes, there is a
large amount of scatter in this scaling law. Also shown are   $3\sigma$ thresholds based on the RMS timing residual
   in the absence of GWs.   The heavy black curves indicate the thresholds  expected when there
   are only white-noise measurement errors.   The light
   (red) curves show thresholds  when red ``timing noise''
    adds to the white noise.  
   Actual noise curves will likely fall somewhere  between the light and heavy curves
   depending on presently uncertain levels of spin noise in MSPs.  
  Figure credit: Jim Cordes.}
\label{fig:rednoise}
\end{figure}

A growing focus of NANOGrav's work is on understanding, characterizing, and exploring mitigation techniques for all sources of noise affecting
pulsar TOAs. 
These sources include those both  extrinsic and intrinsic to the pulsar. Extrinsic sources of noise include radiometer noise, 
 the ISM, and the ionosphere. Intrinsic sources of noise include rotational instabilities and pulse jitter.
Most of these sources of noise are ``white'', with flat power spectral density, and therefore easily distinguished from
the ``red'' spectrum expected due to  background of GWs. However, spin variations from torque fluctuations and internal NS activity display a red spectrum 
 \cite{sc10}, similar to that expected due to GWs, and can have a profound effect on detection prospects (see Figure~\ref{fig:rednoise}). 

Therefore, determining the contribution
of red noise to the residuals is crucial.
Methods used include autocorrelating the residuals, measuring the number of residual zero crossings, and testing the residuals for non-Gaussianity.
The results from these analysis techniques are, so far, similar to those based on spectral analysis presented in the NANOGrav upper limit paper: the majority of NANOGrav pulsars have residuals consistent with white noise, with a few exceptions \cite{dfg+13}.
There are departures from non-Gaussianity but they are consistent with flux modulation due to interstellar scintillation. 
However, as timing precisions increase, it is possible that red spin noise will become apparent in many more pulsars. If red spin noise is shown to be
ubiquitous in MSPs,  GW detection will likely require  
 a large timing program using 50--100 pulsars \cite{cs12}.

Other work involves estimating the contributions of white noise processes to the total timing error budget. For instance,
recent work shows that the RMS for
PSR~J1713+0747 is much greater than expected for radiometer noise alone due to amplitude, profile, and pulse phase variations  \cite{sc12}. Initial efforts to correct for these effects have been unsuccessful, implying that only longer integrations will lead to lower RMS values.
A more detailed discussion of noise processes can be found in the Cordes review in this issue.

In general, we continue to aim for a complete characterization of the noise budget for each NANOGrav pulsar and an understanding
of the ultimate limits to timing precision.
This work may result in some pulsars being removed from the timing program or, perhaps,
observations which are better tuned (for different cadence/frequency/integration time) for a specific pulsar. We are also 
 exploring methods for mitigation of noise processes, with observations of longer-period pulsars providing some hope that
a correlation between timing noise and other observables could lead to mitigation techniques
   \cite{lhk+10}. 

\section{Education and Outreach}
\label{sec:outreach}

An essential component of the NANOGrav mission is to inform the general public about NANOGrav science, inspire the next generation
of scientists, and train student members of NANOGrav to perform research and excel in scientific careers. Our website hosts various
materials that inform the general public. These include descriptions of GWs, pulsars, and GW detection methods, podcasts with NANOGrav astronomers, animations, and several ``Kahn-academy'' style videos which describe our science at a level appropriate for an advanced high-school or beginning undergraduate student. NANOGrav members at all levels are involved in producing these materials.
We also include information on the ``Einstein@Home'' project through which citizen scientists can search for  pulsars in PALFA data 
and on three major outreach projects run by NANOGrav members.  The Arecibo Remote Command Center (ARCC), based at the
University of Texas at Brownsville, and its satellite program at the University of Wisconsin at Madison, involve undergraduate students in pulsar searches.
The Pulsar Search Collaboratory (PSC) is run by West Virginia University and the National Radio Astronomy Observatory and has involved over 800 high-school students from 15 states in pulsar searching. Finally, the
Mid-Atlantic  Relativistic Initiative for Education (MARIE) program, based at Franklin \& Marshall, 
brings NANOGrav students into public high schools 
and hosts astronomy open houses at F\&M.
NANOGrav's goal is to more directly integrate these programs into NANOGrav over the next several years,
building a pipeline for GW studies into the next generation and beyond. We also aim to integrate our efforts
with IPTA-wide outreach.

\section{Partnerships for International Research and Education}
\label{sec:pire}

Since August 2010, many NANOGrav members have received funding through an award from the National Science Foundation's Partnerships for International Research and Education (PIRE) program. The primary goal of PIRE is to ``support high quality projects in which advances in research and education could not occur without international collaboration''. The PIRE funding supports NANOGrav personnel 
 to work on IPTA-related research.
 It also provides funding for IPTA student workshops and science meetings and for research-abroad experiences for U.S. students.  Thus far,  ten NANOGrav undergraduate students and two graduate students
have performed NANOGrav-related research abroad supported by this award. 
Evaluations of the PIRE program demonstrate that the meetings and research abroad experiences are crucial for encouraging students to stay in the field and for preparing them to perform research in an international setting. 
More details about the program, including quarterly evaluation newsletters, can be found at http://nanograv.org/pire.html.

\section{Strategic Planning}
\label{sec:strategic}

\begin{figure}
\includegraphics[scale=0.6]{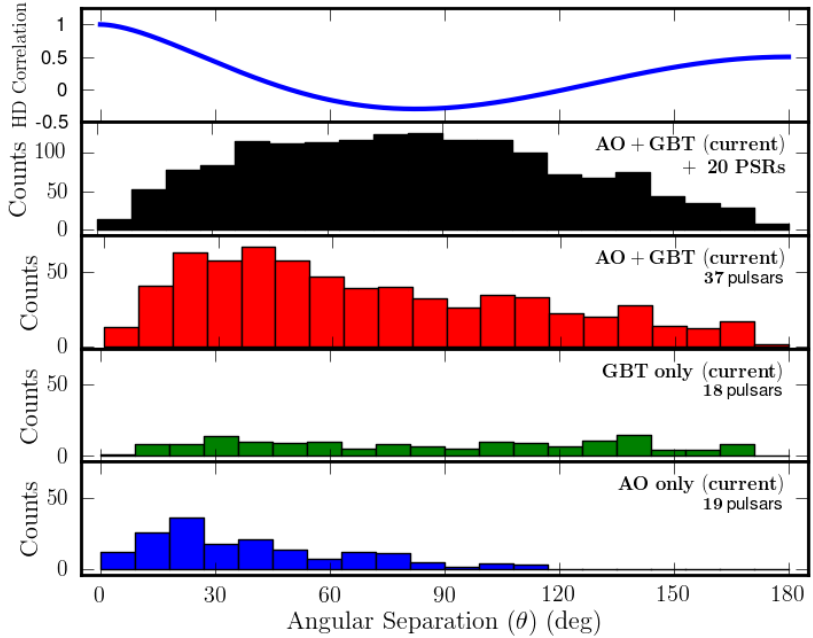}
\caption{
The top panel shows the expected correlation in the timing residuals of
pulsar pairs as a function of angular separation \cite{hd83}. This assumes an isotropic stochastic GW background. The other panels show
the number of pairs as a function of separation for, from bottom to top, MSPs currently timed by NANOGrav  with Arecibo,
 MSPs currently timed by NANOGrav
with the GBT, all MSPs currently timed by
NANOGrav, and all MSPs currently timed
plus an additional 20 uniformly distributed
MSPs. This plot illustrates the dramatically
larger number of pulsar pairs and more complete coverage made possible by the GBT. It
also shows the gains possible if we are able
to add more MSPs to the array through radio pulsar searches in which both the GBT
and AO play critical roles. Note the different y-axis scales on the angular correlation
histograms. Figure credit: Jim Cordes.}
\label{fig:gbtao}
\end{figure}

GW astrophysics was named in the National Academy of Sciences ``New Worlds and New Horizons'' Decadal Report as one of five key discovery areas. Pulsar timing is a critical capability for GW detection and study, as it is the only means to probe  sources in the $10^{-7} - 10^{-9}$~Hz frequency band.
   However, despite  broad support for GW astrophysics, National Science Foundation funding constraints imply uncertain futures for  the GBT and Arecibo, with the GBT recommended for divestment by 2017 and Arecibo operational costs secured for only the next four years. A reduction in time on either of these two telescopes would dramatically impact the sensitivity of our experiment and our time to detection. Carrying out an identical analysis as for NANOGrav's first upper-limit paper but with only Arecibo or only GBT data increases the upper limit by a factor of two. 
Calculating time-to-detection estimates as in the Siemens review article in this issue shows that our sensitivity would be roughly halved and time-to-detection roughly doubled if we had access to only one of these telescopes. Furthermore, Figure~\ref{fig:gbtao} illustrates the importance of both telescopes to a well-sampled
correlation curve and the dramatic increases that will follow from additional pulsars. Therefore, we consider continued access to these telescopes, and the scientific resources required for our observing and analysis programs, to be our most critical strategic planning task.
A secondary goal is to secure funding for ultra-broadband receivers on both Arecibo and the GBT. Receivers with frequency coverage of 700 MHz -- 3~GHz would dramatically increase  sensitivity by boosting overall signal-to-noise ratios and also allowing for more precise
correction of frequency-dependent ISM effects.

\section{Conclusions}
\label{sec:conclusions}

In the five years since its formation, NANOGrav has evolved into  a coherent organization which provides a framework for researches to share ideas and resources and to train students in a collaborative environment. Our work is broad and involves all aspects of GW detection with pulsars from searching for new pulsars to detecting and characterizing different source classes. Over the past several years, our sensitivity has increased through new pulsars and wider bandwidth instruments. We expect
further improvements with new algorithms to increase timing precision, mitigate interstellar scattering, characterize noise, and optimally and efficiently detect and characterize various types of sources. Over the next five years, we will increase our focus on GW sources and multi-wavelength characterization to prepare for the post-detection era of GW astrophysics and build links with the broader astronomy community.

\section{Acknowledgements}

MAM is supported through the NSF PIRE program and the Research Corporation. She also gratefully acknowledges the work of and many discussions with NANOGrav members that have improved this review.
She also thanks the anonymous referees, whose comments significantly improved this paper.

\end{document}